\documentclass[conference]{IEEEtran}
\IEEEoverridecommandlockouts
\usepackage{cite}
\usepackage{amsmath,amssymb,amsfonts}
\usepackage{algorithmic}
\usepackage{graphicx}
\usepackage{textcomp}
\usepackage{xcolor}
\usepackage{xspace}

\usepackage{tabularx}

\usepackage{flushend}

\usepackage{tikz}
\usepackage{nccmath}
\usepackage{soul}
\usepackage{multirow}
\usepackage{graphicx}
\usepackage{array}
\usepackage{listings}
\usepackage{algorithm2e}
\newtheorem{definition}{Definition}

\definecolor{littlegreen}{RGB}{193,236,193}
\definecolor{littlegrey}{RGB}{128,128,128}
\definecolor{altgreen}{RGB}{97,166,46}

    \usepackage{color,colortbl}
    
    \newcommand{\ignore}[1]{}

    \newcommand{\acronym}[1]{{\ensuremath{\sf{#1}}}\xspace}

    \newcommand{\acro}{{\ensuremath{\sf{\mathcal DiCA}}}\xspace}

\newcommand{\dtable}{\acronym{DTable}}

\definecolor{codegreen}{rgb}{0,0.6,0}
\definecolor{codegray}{rgb}{0.5,0.5,0.5}
\definecolor{codepurple}{rgb}{0.58,0,0.82}
\definecolor{backcolour}{rgb}{0.95,0.95,0.92}

\lstdefinestyle{mystyle}{
    commentstyle=\color{codegreen},
    keywordstyle=\color{blue},
    numberstyle=\tiny\color{codegray},
    stringstyle=\color{codepurple},
    basicstyle=\fontsize{9}{9}\ttfamily,
    frame=single,
    breakatwhitespace=false,         
    breaklines=true,                 
    captionpos=b,                    
    keepspaces=true,                 
    numbers=none,                    
    numbersep=4pt,                  
    showspaces=false,                
    showstringspaces=false,
    showtabs=false,                  
    tabsize=2       
}
    
\lstdefinestyle{customasm}{
    belowcaptionskip=1\baselineskip,
    frame=single, 
    frameround=tttt,
    xleftmargin=\parindent,
    language=[x86masm]Assembler,
    basicstyle=\footnotesize\ttfamily,
    commentstyle=\itshape\color{green!60!black},
    tabsize=4,
    numbers=left,
    numbersep=8pt,
    stepnumber=1,
    numberstyle=\tiny\color{gray}, 
    columns = fullflexible,
}

\def\BibTeX{{\rm B\kern-.05em{\sc i\kern-.025em b}\kern-.08em
    T\kern-.1667em\lower.7ex\hbox{E}\kern-.125emX}}

\begin{document}

\title{\huge \acro: A Hardware-Software Co-Design for Differential Check-Pointing in Intermittently Powered Devices}

\author{\IEEEauthorblockN{Antonio Joia Neto, Adam Caulfield, Chistabelle Alvares, Ivan De Oliveira Nunes}
\IEEEauthorblockA{\textit{Rochester Institute of Technology} \\
aj4775@rit.edu,  ac7717@rit.edu, cda5542@rit.edu, ivanoliv@mail.rit.edu}

}

\maketitle

\begin{abstract}

Intermittently powered devices rely on opportunistic energy-harvesting to function, leading to recurrent power interruptions. Therefore, check-pointing techniques are crucial for reliable device operation. Current strategies involve storing snapshots of the device's state at specific intervals or upon events. Time-based check-pointing takes check-points at regular intervals, providing a basic level of fault tolerance. However, frequent check-point generation can lead to excessive/unnecessary energy consumption. Event-based check-pointing, on the other hand, captures the device's state only upon specific trigger events or conditions. While the latter reduces energy usage, accurately detecting trigger events and determining optimal triggers can be challenging. Finally, differential check-pointing selectively stores state changes made since the last check-point, reducing storage and energy requirements for the check-point generation. However, current differential check-pointing strategies rely on software instrumentation, introducing challenges related to the precise tracking of modifications in volatile memory as well as added energy consumption (due to instrumentation overhead).

This paper introduces \acro, a proposal for a hardware/software co-design to create differential check-points in intermittent devices. \acro leverages an affordable hardware module that simplifies the check-pointing process, reducing the check-point generation time and energy consumption. This hardware module continuously monitors volatile memory, efficiently tracking modifications and determining optimal check-point times. To minimize energy waste, the module dynamically estimates the energy required to create and store the check-point based on tracked memory modifications, triggering the check-pointing routine optimally via a non-maskable interrupt. Experimental results show the cost-effectiveness and energy efficiency of \acro, enabling extended application activity cycles in intermittently powered embedded devices.

\end{abstract}

\begin{IEEEkeywords}
Intermittent Computing, Energy Harvesting, Check-pointing
\end{IEEEkeywords}

\section{Introduction}

In contrast to traditional devices that depend on batteries or external power sources, energy harvesting devices capitalize on ambient energy from the surrounding environment to fuel their operations. By leveraging opportunistic energy sources such as solar power\cite{khaligh2017energy} and kinetic energy\cite{beeby2006energy,beeby2015vibration}, intermittent computing enables energy harvesting devices to operate under unpredictable power interruptions. By eliminating the need for batteries, these devices become more sustainable and environmentally friendly, as they reduce electronic waste \cite{tan2011sustainable}. Moreover, battery-less devices offer increased convenience and autonomy since they do not require frequent battery replacements or recharging. This allows for significantly reduced device size and weight, enabling sleeker and more compact designs \cite{kazmierski2011energy}. By removing the need for large batteries, battery-less devices open up new possibilities for miniaturization and integration into various applications \cite{beeby2015vibration}.

On the other hand, power disruptions on intermittent devices present challenges to reliable operation, including data loss, difficulties in state preservation, task resumption, and system instability~\cite{lucia2015simpler,ransford2014nonvolatile}.  As a result, the execution of applications in intermittent devices follows cyclic patterns, where task execution occurs during periods of power availability, followed by power depletion. These cycles require strategies to maintain state across power depletion cycles, ensuring correct operation.

A {\it naive} solution to this problem is the exclusive use of Non-Volatile Memory (NVM), such as FRAM, to store all data. While a completely FRAM-based solution ensures reliability, it increases energy consumption due to increased access latency when compared to volatile memory, such as SRAM. Conversely, an exclusively SRAM-based implementation offers high energy efficiency while lacking reliability across power depletion cycles~\cite{7434963,balsamo2014hibernus}. An alternative approach is to integrate check-pointing into the execution cycles of intermittent devices. Check-pointing involves regularly saving the volatile system state to NVM, creating snapshots that capture the current execution context. Therefore, after power depletion, the system can restore the latest check-point and resume execution properly.

One approach to implement the check-pointing is to modify the original software with additional logic to determine when a minimum energy level has been reached and save the program context to NVM~\cite{ransford2011mementos, bhatti2017harvos, van2016intermittent}. Although functional, these techniques also extend the program's runtime and thus incur additional energy costs, reducing the original application's activity cycle. Alternative techniques propose one-time check-pointing~\cite{balsamo2014hibernus, balsamo2016hibernus} that is triggered when the supplied voltage falls below a certain threshold. However, the latter does not track the modifications made to the volatile memory (VM), requiring the entire VM to be copied to NVM at each check-point. This process also incurs a significant energy cost.

Differential check-pointing is an approach aimed at minimizing the amount of data copied from the VM to the NVM. The fundamental concept is to track modified memory addresses and copy only the modified memory blocks to NVM for each check-point. This approach has led to a notable reduction in both run-time and energy costs associated with check-pointing and has been systematically employed in prior work~\cite{ahmed2019efficient,aouda2014incremental,bhatti2016efficient}. However, prior differential check-pointing schemes are implemented by instrumenting the application source code, still introducing energy and run-time overhead.

\begin{figure}[t]
    \centering
    \includegraphics[width=.45\textwidth]{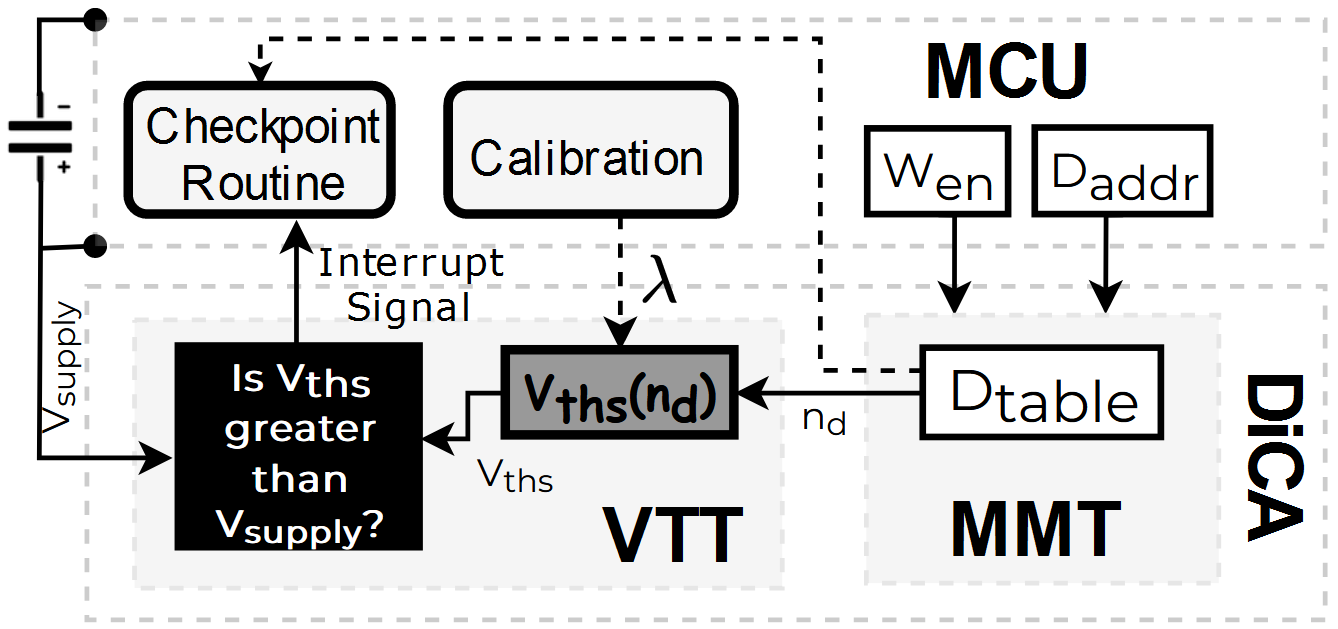}
    \caption{Illustration of \acro's high level architecture}
    \label{fig:overview}
\end{figure}

\subsection{Contributions}

Given the popularity of intermittent computing applications, we argue that future devices could be manufactured with minimal hardware support to facilitate check-pointing and reduce associated energy and run-time costs. With that premise in mind, we propose \acro: a \underline{Di}fferential \underline{C}heck-point \underline{A}ssistant based on a hardware/software co-design. \acro eliminates software instrumentation and application code modifications and dynamically determines optimal differential check-pointing times based on the amount of memory to be saved to NVM. More specifically, \acro comprises:
\begin{itemize}
    \item A lightweight (and energy efficient) hardware module -- called Memory Modification Tracker -- used to efficiently mark modified segments in VM. This module enables tracking of differential memory modifications without requiring any instrumentation or code modification, simplifying and optimizing the check-point generation.
    \item A new interrupt source that optimizes check-point timings and reduces energy consumption. This approach dynamically estimates the minimal supply voltage required to perform the check-point based on the number of segments modified in VM. When the dynamically defined threshold is reached, \acro hardware generates a non-maskable interrupt to initiate the check-pointing procedure.
    \item A software routine that interacts with the \acro hardware to copy appropriate memory segments from VM to NVM based on the optimal parameters configured by \acro hardware module. 
\end{itemize}

The intangibility of applying hardware modifications to real devices has been a significant obstacle for check-pointing techniques, leading to a reliance on software instrumentation. Our approach is rooted in the premise that minimal hardware modifications are both feasible and realistic considering the recent popularity of and demand for energy harvesting devices. We believe that, given their distinct purposes, these devices can benefit from simple, practical, and inexpensive hardware modifications to enhance their performance without requiring massive architectural overhauls. 

\subsection{Scope}

Battery-less and intermittent computing devices are typically implemented with micro-controller units (MCUs) that run software at bare-metal, have low-cost, and are energy efficient. In this work, we focus on ultra low-energy MCUs (e.g., Atmel AVR ATMega~\cite{Microchip_2015}, TI MSP430~\cite{TI-MSP430}) which feature single-core $8$- or $16$-bit CPUs running at low clock frequencies (usually $1$ to $16$ Mhz). They use between $4$ and $16$ KBytes of SRAM as VM while the rest of the address-space is available for NVM. We implement \acro prototype atop an open-source version of TI MSP430 from openCores~\cite{openmsp430}.

\subsection{Organization}

This paper is structured as follows. Section~\ref{sec:overview} presents the high-level ideas in \acro design. Section~\ref{sec:details} delves into the details of \acro architecture and specifies \acro hardware and software components. Section~\ref{sec:eval} discusses the implementation of \acro prototype, the experimental set-up, and presents \acro empirical evaluation. Section~\ref{sec:rw} discusses related work and Section~\ref{sec:conclusion} concludes the paper.

\section{\acro High-Level Overview}\label{sec:overview}

\acro is a hardware/software co-design. It includes an inexpensive hardware module that tracks modified VM segments. 
Compared to methods that perform this tracking in software, it reduces energy consumption, enabling more instructions belonging to the original application to be executed per power cycle. \acro hardware also implements a new interrupt source that triggers the check-point generation, i.e., the process of copying modified VM segments to NVM, based on the estimated required time and available energy. Figure~\ref{fig:overview} illustrates \acro architecture.

\begin{figure}
    \centering
    \includegraphics[width=.45\textwidth]{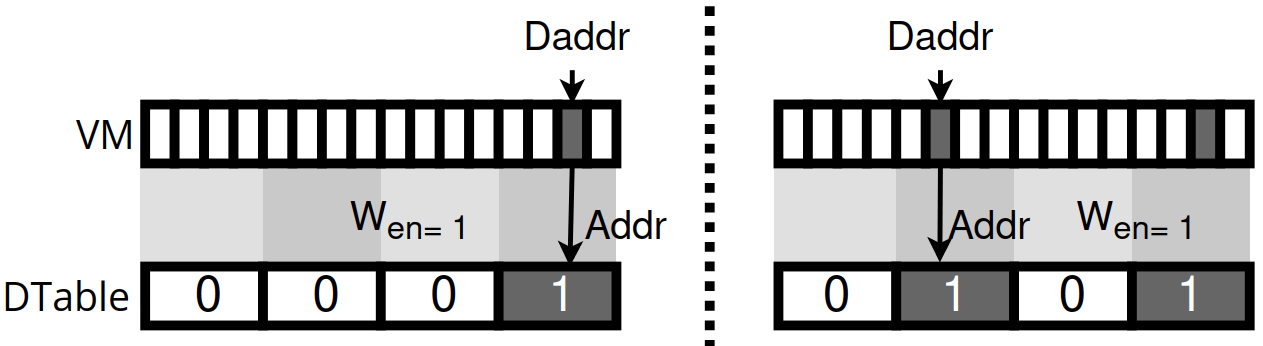}
    \caption{Illustration of \dtable memory tracking}
    \label{fig:dtanle}
\end{figure}

\subsection{\acro Hardware}

\acro hardware has two sub-modules: Memory Modification Tracker (\acronym{MMT}) and Voltage Threshold Tracker (\acronym{VTT}).

\acronym{MMT} divides the volatile memory into blocks. It detects whether these blocks have been written to by monitoring two internal CPU signals: the write enable bit (denoted $W_{en}$), which indicates whether the MCU is writing to the memory, and the $D_{addr}$ signal, that defines the memory address being written to when $W_{en}=1$. Whenever $W_{en}$ is active, \acro takes the value of $D_{addr}$ and uses it as an index to set a bit in a register vector called (Dirty-bits Table) \dtable, indicating that the block to which the address $D_{addr}$ belongs to has been modified.
Figure~\ref{fig:dtanle} depicts \acro updating \dtable as VM blocks are modified.
To track the differential changes, \acro also has a controller that allows software to reset \dtable after loading the prior check-point.

(\acronym{VTT}) generates an interrupt signal based on how many memory segments in VM have been modified (as detected by \acronym{MMT}). It dynamically calculates a threshold voltage supply value, denoted $V_{ths}$. The value of $V_{ths}$ is determined by counting the number of modified VM memory blocks, i.e., the number of active bits in \dtable, to allow sufficient time for the check-pointing routine. When the supplied voltage falls below $V_{ths}$, indicating an impending power depletion, a non-maskable interrupt is triggered.

\subsection{\acro Software}\label{sec:dica_sw_overview}

\acro software component is implemented as an interrupt service routine (ISR) associated with the \acronym{VTT}-generated interrupt. Based on \dtable, the ISR captures a snapshot of modified segments in VM and CPU registers, and copies them to NVM.
After the successful check-point generation, the system is powered off until the energy harvesting component recharges the power supply. When the supplied voltage reaches a full charge threshold (denoted $V_{full}$), the MCU restarts and \acro software restores the most recent check-point, allowing the system to resume operation from the suspended state.








\section{\acro in Details}\label{sec:details}
This section details \acro design. For quick reference, Table~\ref{tab:notation} summarizes the notation used in the rest of the paper.
\vspace{-2em}
\begin{table}[h]
    \centering
    \caption{Notation Summary}
    \label{tab:notation}
    \resizebox{\columnwidth}{!}{%
    \begin{tabular}{lp{5.8cm}}
    \hline
    \textbf{Notation} & \textbf{Description}                                                   \\ \hline 
    \rowcolor[HTML]{EFEFEF} \dtable &   Bit-vector indicating memory blocks that have been modified since the last check-point\\
     $\dtable_{\acronym{size}}$ & Size of \dtable (in number of bits) \\
     \rowcolor[HTML]{EFEFEF} \dtable[d] &  d-th bit in \dtable  \\
     $n_d$ & Number of active bits in \dtable \\
     \rowcolor[HTML]{EFEFEF} $reset$ & A 1-bit CPU signal that sets all the DTable bits to zero\\
     $W_{en}$  & A 1-bit signal that indicates if the MCU is writing to memory \\
     \rowcolor[HTML]{EFEFEF}$D_{addr}$ & CPU signal containing memory address being modified by the MCU when $W_{en}=1$\\
    $BSS$ & Block size shift\\
    \rowcolor[HTML]{EFEFEF} $SP$ & Stack Pointer\\
    $SP_{Lim}$ & Stack Pointer Limit: lowest memory address for data allocation in the CPU stack architecture\\
    \rowcolor[HTML]{EFEFEF}\acronym{NVM} and\acronym{VM} & Non Volatile Memory and Volatile Memory\\
    $\acronym{VM}_{\acronym{size}}$ & Volatile Memory size\\
    \rowcolor[HTML]{EFEFEF} $\acronym{VM}_{min}$,$\acronym{VM}_{max}$ & Maximum and Minimum memory addresses of \acronym{VM} \\
     $SF_{mask}$ & Stack Frame mask \\
     \rowcolor[HTML]{EFEFEF} $\acronym{VM}^{block}_{size}$ & Size of a memory block \\
     $V_{ths}$ & Supply voltage threshold \\
     \rowcolor[HTML]{EFEFEF} $V_{ths}(n_d)$ & Function mapping $n_d$ to $V_{ths}$  \\
    \hline
    \end{tabular}%
    }
\end{table}

\subsection{Memory Modification Tracker (MMT)}

\acronym{MMT} is a hardware sub-mudule designed, in its simplest form, to monitor the memory locations written by the CPU. Central to this component is the \dtable, a peripheral 
that enables efficient differential check-point generation, obviating the need for instrumentation or application-specific code modifications.

\dtable is a bit-vector where each bit maps to a memory block, sequentially, in \acronym{VM}. Memory blocks are of pre-defined size denoted as $\acronym{VM}^{block}_{size}$.
\dtable's size is determined by the total size of the \acronym{VM} ($\acronym{VM}{\acronym{size}}$) divided by $\acronym{VM}^{block}_{size}$.
$\acronym{VM}^{block}_{size}$ defines the granularity of memory tracking and is a design parameter chosen at MCU manufacturing time. To simplify the hardware implementation, we restrict $\acronym{VM}^{block}_{size}$ to powers of 2.

\acronym{MMT} monitors the CPU signals $W_{en}$ and $D_{addr}$ which are used by the CPU to write to memory.
As part of the underlying CPU behavior, $W_{en}$ is set to $1$ whenever a write access to memory occurs, whereas $D_{addr}$ contains the address of the memory location being written when $W_{en}=1$. Therefore, whenever $D_{addr}$ is within \acronym{VM} and $W_{en}=1$, \acronym{MMT} determines a \dtable index ($Addr$) by shifting the relative address of \acronym{VM} ($D_{addr}-\acronym{VM}_{min}$) by \acronym{BSS} bits, 
where $\acronym{BSS} = \log_{2}^{\acronym{VM}^{block}_{size}}$.
Then \acronym{MMT} sets the bit corresponding to $Addr$ in \dtable to $1$. 
Figure~\ref{fig:dtanle} illustrates \dtable functionality. The specification of \acronym{MMT} basic behavior is presented in Definition~\ref{def:MMT_basic} (see Section~\ref{sec:MMT_extended} for \acronym{MMT} extended version that ignores de-allocated stack frames during check-point generation).

\vspace*{1em}
\begin{tabular}{|p{0.9\linewidth}|}\hline 
\begin{definition}\label{def:MMT_basic}
    {Memory Modification Tracker Model}

    \begin{equation*}
        i \in [ 1, \dtable_{\acronym{size}}] 
    \end{equation*}
    
    \begin{equation*}
        Addr := (D_{addr}-\acronym{VM}_{min}) \gg BSS
    \end{equation*}

    \[ \dtable [i] :=
  \begin{cases}
    0      & \quad \text{if} \quad reset \\
    1      & \quad \text{if} \quad  (i=Addr)\quad \land  \\
    & \quad  (D_{addr}\in \acronym{VM}) \land W_{en} \\ 
    \dtable [i]      & \quad \text{Otherwise} \\
    
  \end{cases}
\]
    
 \vspace*{.1cm}
\end{definition}\\\hline
\end{tabular}
\vspace*{1em}

\begin{figure}
    \centering
    \includegraphics[width=\columnwidth]{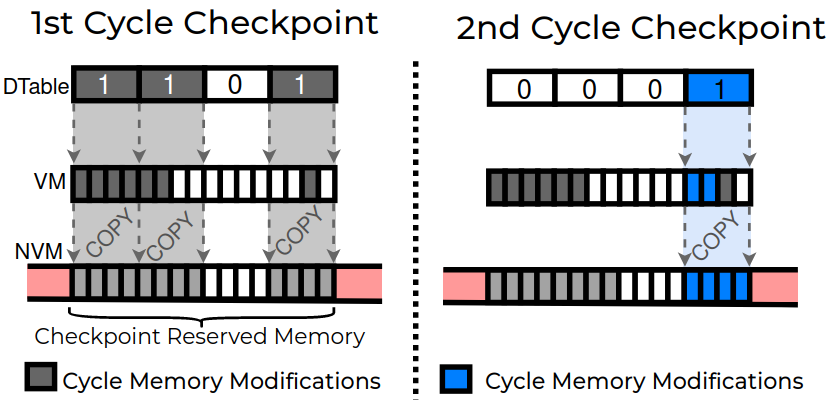}
    \caption{Visualization of differential check-pointing using \dtable}
    \label{fig:diff check-point}
\end{figure}

\textbf{Rationale}. 
\acronym{MMT} detects differential memory changes between the last check-point and the current memory state. This feature reduces the number of memory blocks that must be copied to NVM in the next check-point, as unmodified data remains consistent in NVM and need not be copied.
To support this functionality, the \acronym{MMT} module incorporates a control bit ($reset$) that clears \dtable on each power cycle (i.e., at MCU boot). Figure~\ref{fig:diff check-point} illustrates the differential check-pointing process across subsequent check-points based on \dtable.

    \subsection{Extending \acronym{MMT} to Ignore De-Allocated Stack Frames}\label{sec:MMT_extended}

    \acronym{MMT} basic design only sets \dtable bits to track memory modifications.
    However, it does not clear \dtable if a memory block is no longer in use by the program. Since functions are called (and returned from) multiple times in most programs, \acronym{MMT} basic design would check-point function stack frames that are no longer in use. Considering nested function calls, this check-pointing approach would unnecessarily include a large number of memory blocks related to stack frames that are no longer in use. To address this issue, \acronym{MMT} is extended to clear \dtable bits associated with stack frames of functions upon their completion.

\begin{figure}[t]
    \centering
    \includegraphics[width=.9\columnwidth]{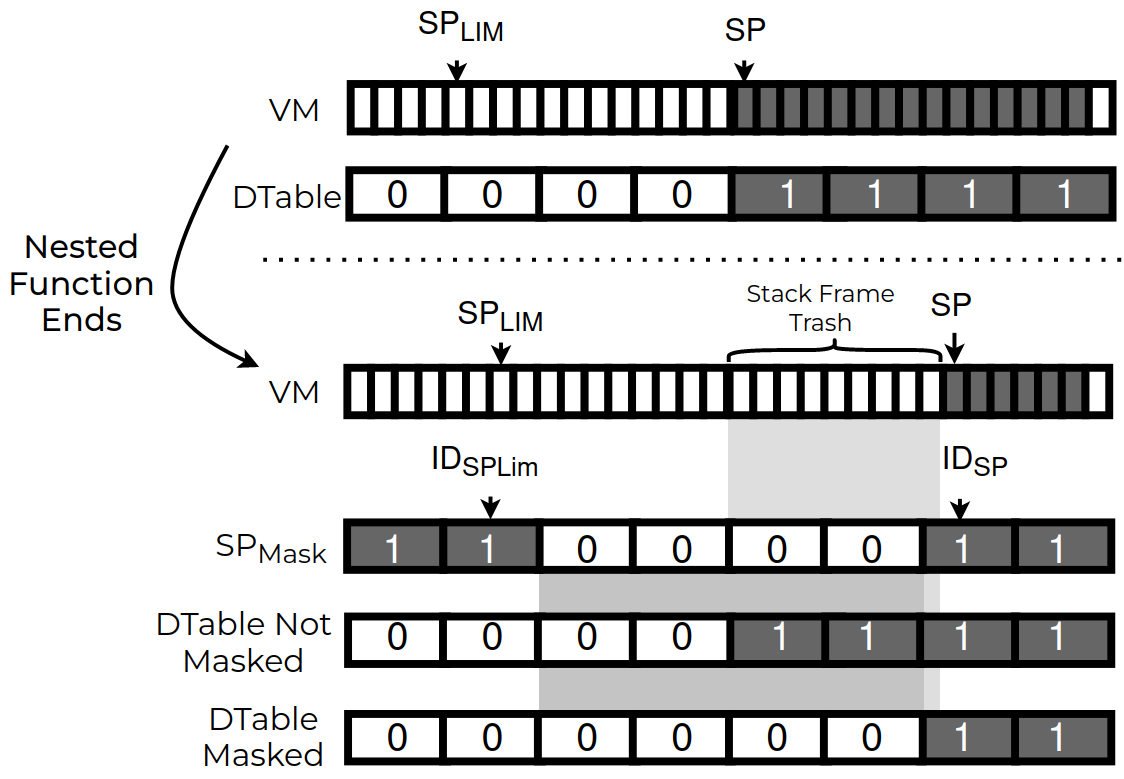}
    \caption{Visualization of the \dtable stack frame trash cleaning }
    \label{fig:stack_clean}
\end{figure}

The stack frame cleaning is depicted in Figure~\ref{fig:stack_clean}. It uses a bit mask, called $SF_{mask}$, that masks \dtable entries. This mask indicates whether the corresponding D-table bit is no longer in use. $SF_{mask}$ values are determined based on two values. The first, $SP_{LIM}$, defines the lowest memory address\footnote{The use of the ``lowest memory address'' as the limit assumes that the stack grows downwards in the underlying MCU architecture.} that can be used for stack allocation in the MCU. The second, $SP$, is the current stack pointer, i.e., a CPU signal that contains the address of the top of the currently allocated stack. Therefore, at any given time, the memory region between $SP$ and $SP_{LIM}$ is unallocated.

To produce $SF_{mask}$, the relative memory positions of $SP$ and $SP_{LIM}$ with respect to the \acronym{VM} are computed by subtracting $\acronym{VM}_{min}$ from both input parameters. $\acronym{VM}_{min}$ indicates the lowest memory address of \acronym{VM}. Next, the indices that correspond to these relative memory addresses in \dtable are obtained by shifting the resulting values by $BSS$. This generates the indices $ID_{SP}$ and  $ID_{SP_{LIM}}$. The mask is then generated by setting all indices between $ID_{SP}$ and $ID_{SP_{LIM}}$ to 0 and leaving all others as 1. \acronym{MMT} behavior, extended with the stack frame cleaner, is specified in Definition~\ref{def:MMT_extended}.\\

\begin{figure}
    \begin{tabular}{|p{0.9\linewidth}|}\hline 
        \begin{definition}\label{def:MMT_extended}
            {Memory Modification Tracker with Stack Frame Cleaner}:        
        
    \begin{itemize}
        \item Stack Frame mask:
    \end{itemize}

    \begin{equation*}
        i \in [ 1, \dtable_{\acronym{size}}] 
    \end{equation*}

            \begin{equation*}
                ID_{SP} := ((SP - \acronym{VM}_{min}) \gg BSS)
            \end{equation*}

            \begin{equation*}
                ID_{SP_{LIM}} := ((SP_{Lim}- \acronym{VM}_{min}) \gg  BSS) 
            \end{equation*}

            \begin{equation*}
                SF_{mask}[i] := (i \leq ID_{SP})  \lor   (i \geq ID_{SP_{LIM}}) 
            \end{equation*}

\begin{itemize}
    \item \dtable with Stack Frame Cleaner:
\end{itemize}
    



    \[ \dtable [i] :=
  \begin{cases}
    0      & \quad \text{if} \quad SF_{mask}[i] \lor reset \\
    1      & \quad \text{if} \quad  (i=Addr)\quad \land  \\
    & \quad  (D_{addr}\in \acronym{VM}) \land W_{en}   \\ 
    
    \dtable [i]      & \quad \text{Otherwise} \\
    
  \end{cases}
\]
\end{definition}~\\\hline
\end{tabular}
\end{figure}

\subsection{ Voltage Threshold Tracker }

To reduce power consumption, \acro establishes the ideal moment to start the check-pointing routine.
To this end, it implements a non-maskable interrupt source that is triggered when the system's supplied voltage ($V_{supply}$) falls below a dynamically defined threshold ($V_{ths}$). $V_{ths}$ serves as a proxy for the amount of energy required for the check-pointing routine to fully execute before $V_{supply}$ drops below to a level that is insufficient to sustain MCU operation. $V_{ths}$ is calibrated based on the number ($n_d$) of active bits of in \dtable (see Section~\ref{sec:calibration} for details). When $V_{ths}$ is reached, the interrupt is triggered.

$n_d$ default value is $0$ after loading a check-point. It is incremented by one whenever a \dtable value changes from 0 to 1 (as detected with the operation $[\neg \dtable[Addr] \land W_{en}]$). Conversely, when a bit in the \dtable is cleared (due to stack frame cleaning) $n_d$ is decremented. To determine the number of bits cleared due to stack frame cleaning, \acro checks the previous stack pointer index, denoted $ID_{SP}^{t-1}$ and subtracts it from the current index $ID_{SP}^t$. If the result is greater than zero, it is subtracted from $n_d$. The value of $n_d$ is then used to compute the $V_{ths}$, as detailed in Section~\ref{sec:calibration}. 
The interrupt signal generation is specified in Definition~\ref{def:interrupt}.

\vspace*{1em}
\begin{tabular}{|p{0.9\linewidth}|}\hline 
\begin{definition}\label{def:interrupt}
    {Voltage Threshold Tracker Model}:

    \begin{itemize}
        \item Stack Frame reduction counter:
    \end{itemize}

    \begin{equation*}
            ID_d := ID_{SP}^t - ID_{SP}^{t-1}
    \end{equation*}

    \begin{itemize}
        \item Dtable bit counter:
    \end{itemize}

    \[ n_d :=
  \begin{cases}
    0       & \quad \text{if } reset\\
    n_d+1  & \quad \text{elif } (\neg \dtable[Addr] \land W_{en}) \\
    n_d-ID_d  & \quad \text{elif } (ID_d > 0) \\
    n_d  & \quad \text{Otherwise} 
  \end{cases}
\]

   \begin{itemize}
        \item Interrupt Signal:
    \end{itemize}

    \begin{equation*}
            IT_{sig} \leftarrow  V_{supply} < V_{ths}(n_d)
    \end{equation*}
    
\end{definition}\\\hline
\end{tabular}

\subsection{Voltage Threshold ($V_{ths}$) Calibration}\label{sec:calibration}

$V_{ths}$ is determined by two factors: $n_d$ and a constant $\lambda$. A device calibration phase is conducted at system deployment time to determine $\lambda$. The calibration process involves fine-tuning $\lambda$ to the specific MCU in order to obtain an optimal $V_{ths}(n_d)$ function for that particular device.
Similar to prior work~\cite{balsamo2016hibernus}, we assume that the voltage supply decay between $3.6$V (fully charged supply) to $2.0$V (minimal operational threshold) is linear. Therefore, $\lambda$ is determined by measuring the total amount of blocks that can be copied in one full device power cycle ($N$) and dividing $1.6$V by $N$.

\textbf{Implementation of $V_{ths}(n_d)$}: In order to avoid hardware multiplications, the product of $\lambda$ and $n_d$ is computed using addition operations. Initially, the threshold voltage $V_{ths}$ is set to $V_{min}$. As $n_d$ increases or decreases, an additional register $n_d'$ tracks the values of $n_d$ with a unitary increment or decrement. This additional register is necessary because $n_d$ can decrease by more than one value at a time, due to the stack frame cleaner. 
This is specified in Definition~\ref{def:vth}.

    \vspace*{1em}
    \begin{tabular}{|p{0.9\linewidth}|}\hline 
        \begin{definition}\label{def:vth}
            {$V_{ths}$ Calibration Model}:        

\begin{equation*}
    V_{ths}(0) = V_{min}
\end{equation*}

  \[ n_d' :=
          \begin{cases}
            n_d' + 1        & \quad \text{if} \quad n_d > n_d' \\
            n_d' - 1        & \quad \text{if} \quad  n_d < n_d'  \\
            n_d'             &  \quad \text{if} \quad n_d = n_d' \\
  \end{cases}
\]
                
            \[ V_{ths} :=
          \begin{cases}
            V_{ths} + \lambda     & \quad \text{if} \quad n_d > n_d' \\
            V_{ths} - \lambda       & \quad \text{if} \quad  n_d < n_d'  \\
            V_{ths}     & \quad \text{if} \quad n_d = n_d'  \\
    
  \end{cases}
\]

\vspace*{0mm}
\end{definition}\\\hline
\end{tabular}
\vspace*{1em}
\subsection{Check-Point Generation}

When the check-pointing ISR is triggered ($IT_{sig} = 1$), \acro software executes to copy \acronym{VM} memory blocks that are marked in \dtable to a dedicated region in NVM. Enough space in \acronym{NVM} should be reserved for this purpose.

Before the check-pointing process, a flag stored in the NVM is set to $True$, indicating the active state of the check-pointing process. Once check-pointing is completed, the flag is unset. If the check-pointing process is not successfully completed, the value of $\lambda$ can be adjusted to a more conservative value for the subsequent power cycle.

In order to generate the VM snapshot using the \dtable, the \acro software iterates through each bit of \dtable. If a bit is 0, the iteration proceeds to the next bit. If the bit is 1, the memory block associated with that bit (of size $\acronym{VM}^{block}_{size}$) is copied to its corresponding position in NVM. This process is shown in Algorithm 1.

\SetKwComment{Comment}{/* }{ */}
\RestyleAlgo{ruled} 
\SetKw{KwBy}{by}
\begin{algorithm}
\caption{Memory check-pointing using \dtable}\label{alg:two}
\KwData{\dtable,$\dtable_{\acronym{size}}$; $\overline{\acronym{NVM}}$ = pointer to the first position of nonvolatile memory related to the check-point, $\acronym{VM}$ = pointer to the first position of the ram memory, $b$ = $\acronym{VM}^{block}_{size}$}

\For{$i\gets0$ \KwTo $\dtable_{\acronym{size}}-1$}{
    \If{$\dtable[i]$ is 1}{
        memcpy($\overline{\acronym{NVM}}[b*i]$, $\acronym{VM}[b*i]$, $b$);
        }
}
\end{algorithm}

\subsection{Execution Resumption}

When resuming execution in a new power cycle, $V_{ths}$ can be re-calibrated by adjusting $\lambda$. The system must also check the integrity of the current check-point. If no check-point is found or if it is corrupted (or incomplete), the application starts anew. If a valid check-point exists, it must be reloaded by copying the check-point data from NVM to the VM, and restoring the CPU registers. Additionally, after copying the check-point data to NVM, \dtable and $n_d$ are cleared. This step sets \acro up for check-pointing the next power cycle. Finally, the Program Counter (PC) is loaded with the address that was meant to be executed by the application in the previous power cycle immediately before \acro interrupt was triggered.

\section{Prototype \& Experiments}\label{sec:eval}

We synthesize \acro using a Xilinx Artix-7 FPGA~\cite{artix7}, on a Basys-3~\cite{basys3} prototyping board. The Xilinx Vivado tool-set~\cite{vivado} was used for synthesizing the on top of the openMSP430~\cite{openmsp430} MCU core. \acro was written in the Verilog hardware description language. Each module implements the logic outlined in Section~\ref{sec:details}. \acro was implemented in 368 lines of Verilog code for the MMT and VTT hardware modules (including their integration with the underlying openMSP430 core) and 201 lines of C code for \acro software ISR and check-point recover. 

In addition to the FPGA-based prototype, we perform complementary experiments using a low-energy device for realistic energy results. We use the MSP430FR2476 MCU for these experiments, featuring  8kB of SRAM (VM) and 64KB of FRAM (NVM) and running at a CPU clock frequency of 1MHz.  

\subsection{Profiling $\acronym{VM}^{block}_{size}$ }\label{subsec:block_profiling}

\begin{figure}
    \centering
    \includegraphics[width=.9\columnwidth]{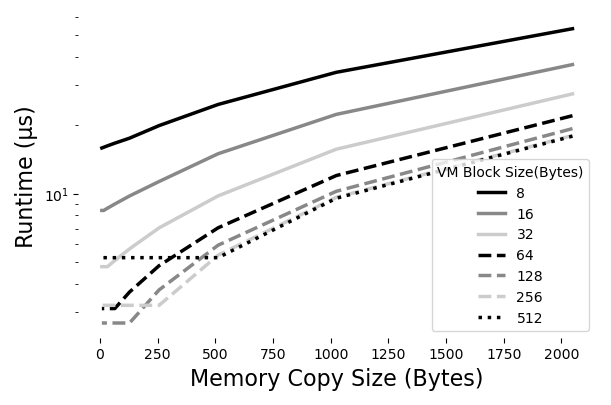}
    \vspace{-1em}
    \caption{Profiling of copy run-time vs. $\acronym{VM}^{block}_{size}$}
    \vspace{-1em}
    \label{fig:dtableprofile}
\end{figure}

An important manufacturing time decision is to determine $\acronym{VM}^{block}_{size}$ in \acro. This decision has a direct impact on the number of bits in \dtable, thereby affecting the hardware size. However, it also plays a vital role in enhancing the granularity of memory-tracking, which has the potential to reduce check-pointing time and energy consumption. To determine the optimal value of $\acronym{VM}^{block}_{size}$, we profile the check-pointing run-time against various values.
The results for $\acronym{VM}^{block}_{size}$ varying from 8 to 512 Bytes are presented in the Figure~\ref{fig:dtableprofile}.

The results show that smaller $\acronym{VM}^{block}_{size}$ incur longer run-times for copying the same amount of memory, even for small memory amounts. This can be attributed to the inherent overhead associated with the copy operation, encompassing memory address calculations, data transfers, and synchronization. Moreover, smaller $\acronym{VM}^{block}_{size}$ result in larger \dtable sizes, necessitating a more extensive search to identify modified bits, which further contributes to the augmented check-pointing run-time. Conversely, as the $\acronym{VM}^{block}_{size}$ increases, it becomes apparent that larger values are constrained by a minimum overhead, which increases proportionally due to the granularity of memory-tracking.

Based on these experiments, we determined that the optimal $\acronym{VM}^{block}_{size}$ for the MSP430FR2476 is $128$ Bytes, as it provides the most favorable balance between run-time and granularity. When selecting such parameters, it is crucial to take into account the characteristics of the device's memory and its clock frequency. These factors can vary across different devices and significantly influence the profile. Consequently, it is important to highlight that our prototype's profile may not be optimal to other devices or MCU models. That being said, we use $128$-Byte blocks as the reference value for the remainder of the experiments in \acro evaluation.

\subsection{Hardware Footprint Overhead}

\begin{figure}
    \centering
    \includegraphics[width=\columnwidth]{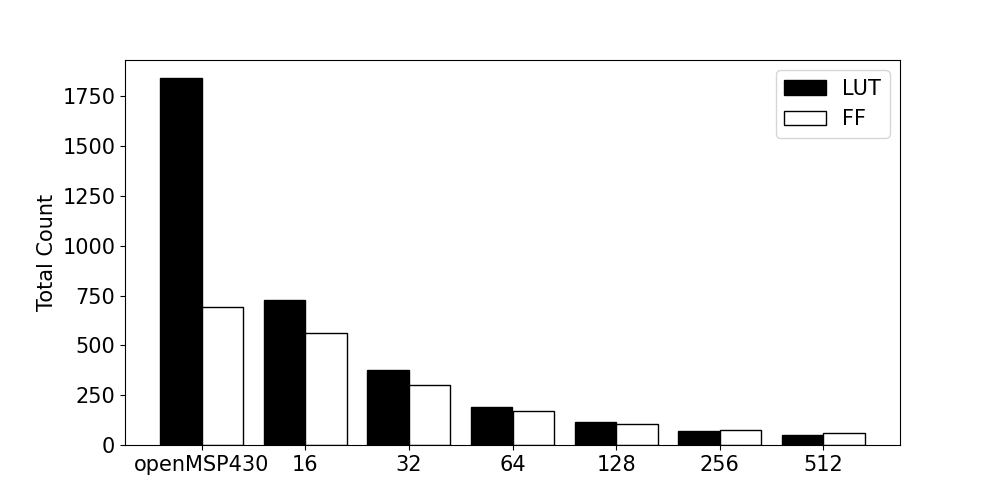}
    \vspace{-2em}
    \caption{Hardware cost of unmodified openMSP430 compared to additional cost of \acro with $\acronym{VM}^{block}_{size}$ of 16 to 512 Bytes}
    \vspace{-1em}
    \label{fig:hardware}
\end{figure}

We assess the hardware cost in terms of additional Look-up Tables (LUTs) and flip-flops/registers (FFs). The increase in LUTs reflects the additional chip cost/size attributed to combinatorial logic. The increase in FFs indicates the additional state required by sequential logic. Figure~\ref{fig:hardware} shows the hardware cost of unmodified openMSP430 and the additional cost of \acro hardware when configured to monitor memory blocks from 16 to 512 Bytes.

The cost of \acro hardware is maximized when $\acronym{VM}^{block}_{size}$ is at the lowest value of 16-Bytes. With this configuration, \acro hardware incurs the maximum overhead with additional 730 LUTs and 561 FFs. However, this additional cost decreases as $\acronym{VM}^{block}_{size}$ increases. For instance, configuring \acro hardware to monitor 512-Byte blocks only requires 49 LUTs and 58 FFs. For a 128-Byte configuration, which we determined has an ideal check-pointing runtime (see Section~\ref{subsec:block_profiling}), additional 114 LUTs and 106 FFs are required. Configuring \acro hardware to monitor 128-Byte blocks causes an increase of $\approx8.7$\% relative to the unmodified openMSP430 core.

\subsection{Hardware Energy Overhead}

While the added hardware modules eliminate the need for software-based memory tracking (and associated energy consumption), they also drain energy at runtime.
We use Vivado synthesis tool to estimate \acro's power consumption on the Basys3 FPGA board. In this analysis, we consider \acro configured for 128-Byte memory blocks.
The MCU, including openMSP430 default set of peripherals and \acro, consumes 89 mW of static power whereas \acro hardware alone is reported\footnote{Vivado does not report energy consumption units under 1mW, treating such small values as negligible. Thus, 1mW is the upper bound for \acro consumption because it is not possible to obtain precise measures below this value. In reality, however, \acro may consume significantly less than $1$ mW.} to draw less than $1$ mW. Therefore, \acro is responsible for less than $1.1$\%, of the device's static power consumption. 
%
%

The dynamic power drawn depends on the frequency of memory writes performed by the software. We consider an application that writes to all memory blocks in a loop to evaluate the worst case. In this scenario, the unmodified openMSP430 (along with peripherals) draws 234 mW of dynamic power. When equipped with \acro, the dynamic power increases to 241 mW, representing a $\mathcal{\approx}$ 2.9\% increase.

We note that these estimates are based on the FPGA deployment and may vary once the MCU design is manufactured as an integrated circuit.




\begin{figure*}
    \centering
    \includegraphics[width=\textwidth]{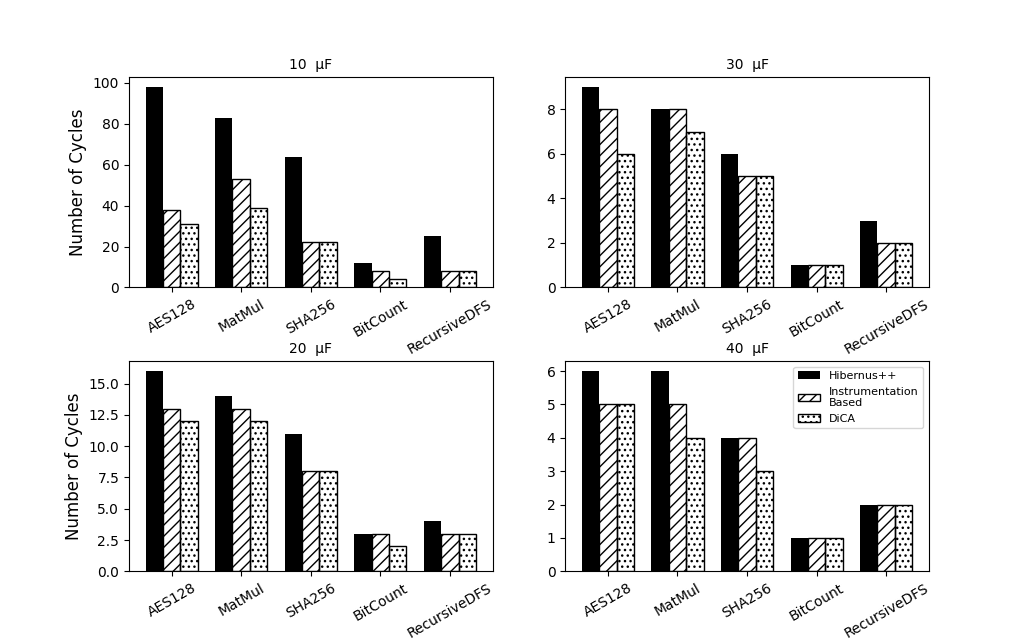}
    \caption{Number of power cycles required to compute five application test-cases: \acro vs. two related check-pointing methods}
    \label{fig:results}
\end{figure*}

\subsection{Power Cycle Efficiency}

To evaluate the efficacy of \acro, we compare its performance with prior related work with respect to the amount of power cycles required to complete five distinct computations. Our benchmark considers five well-known algorithms: an AES encryption block cypher, a matrix multiplication, a SHA256 cryptographic hash function, a bit counting function, and a Depth-First Search (DFS) algorithm. To gauge how well our approach performs in comparison to other methods that trigger the check-point based on Voltage thresholds, we implemented an optimized version of Hibernus++~\cite{balsamo2016hibernus} (see related work in Section~\ref{sec:rw}). Our implementation of Hibernus++ triggers the check-point within the time required to copy VM (8 kB) before energy depletes. Furthermore, to compare our method with existing differential check-pointing approaches, we integrated a software instrumentation version of a memory modification monitor into Hibernus++, drawing upon prior work~\cite{dice} as a reference.

In our experiment, we set up four distinct experimental configurations, each considering an MCU equipped with a capacitors of different capacitance, used to store harvested energy. By varying the capacitance values, our objective is to examine the influence of varying power supply decays on the performance of \acro. The results are presented in Figure~\ref{fig:results}.

 Across the four experimental capacitance configurations, we observed that \acro consistently outperforms Hibernus++. It also outperforms the instrumentation-based differential check-pointing in most cases, requiring fewer execution cycles to run the bench-marked algorithms. The impact of this outcome becomes more pronounced in setups characterized by lower capacitance, where the rate of power supply decays is elevated. This occurs because, with more power cycles, the check-pointing routine tends to occupy a more substantial portion of the MCU execution time. As a result, the difference prior methods and \acro is more pronounced.

It's important to note that \acro and other techniques from the literature are not mutually exclusive. \acro is compatible and interoperable with a diverse array of existing strategies, allowing for a potential combination of approaches to further enhance the efficiency and effectiveness of inttermittent applications.

 

\section{Related Work}\label{sec:rw}


\subsection{Traditional Check-Pointing}
Differential (a.k.a. incremental) check-pointing~\cite{narayanasamy2005bugnet, randell1978reliability, plank1994libckpt, agarwal2004adaptive, koo1987checkpointing} is widely studied for traditional computing systems. These systems are developed to operate on high-end devices and thus prioritize performance. Since they do not consider devices that operate on intermittent power supplies or have limited computational/storage resources, these techniques rely on tasks that low-end MCUs are not capable of performing, such as maintaining complex data structures~\cite{kim2022listdb, agarwal2004adaptive}, or relatively expensive hardware support~\cite{egger2016efficient, park2011fast, plank1994libckpt} to compute the differentials and store the check-point. Check-points in these systems also have different purposes, e.g., fault tolerance, load-balancing, and data concurrency across parallel servers. 

\subsection{Check-Pointing in Intermittently Powered Devices}
Energy-efficient check-pointing schemes are required for intermittently powered devices that harvest their own supply of energy. Compile-time techniques~\cite{ransford2011mementos, bhatti2017harvos} instrument the application source code with additional function calls that check the device's current power. They are placed at specific points in the program's control flow, such as each backward edge of a loop or function returns.
To properly create a check-point, the program context must be written in NVM, including all registers and data currently in use. Therefore frequent check-pointing can result in additional energy losses.

To reduce the overhead, runtime techniques~\cite{balsamo2014hibernus, balsamo2016hibernus} determine the proper time for a check-point to be made while the device operates. The time to create a check-point is determined by employing a hardware-based interrupt to check the voltage supply periodically, and a check-pointing routine is executed once the voltage reaches a minimum threshold, reducing the number of check-points. In contrast with \acro, prior runtime techniques save all registers and the entire VM at each check-point. 

An alternative approach to further reduce the size of check-points is to save just the changes in a similar manner to differential check-points in traditional computing. However, computing the differentials in an energy-efficient and low-cost manner for intermittent computing devices is challenging. Some techniques use software to iterate over all addresses of VM~\cite{bhatti2016efficient} and compare them to the prior check-point. Others compare the hashes of memory blocks~\cite{aouda2014incremental}. DICE~\cite{dice} computes the differentials by maintaining an internal \textit{modification record}, and the application software is instrumented with function calls that records differentials. DINO~\cite{lucia2015simpler} extends C's programming model by providing compiler-aided analysis to place \textit{task-boundaries}, where check-points and data versioning takes place. As an alternative, new programming abstractions that operate on data in NVM at all times have been proposed. For instance, Chain~\cite{colin2016chain} proposes a new program abstraction that guarantees that the state of self-contained tasks is preserved in NVM. Unlike \acro, these techniques either depend on additional software, software modification via instrumentation, or new programming abstractions.

\section{Conclusion}\label{sec:conclusion}

We proposed \acro, a lightweight hardware/software co-design to support efficient differential check-pointing for intermittently powered devices. \acro eliminates the need for application code modifications or instrumentation, simplifying and optimizing check-point generation. In addition, it implements a non-maskable interrupt to dynamically estimate optimal check-pointing times, thereby increasing the active period of applications during a power cycle. \acro interrupt triggers a software routine that complements the hardware, efficiently copying modified memory segments from volatile to non-volatile memory.
We implemented and evaluated \acro with an FPGA deployment. \acro open-source prototype is available at~\cite{repo}.

\section*{Acknowledgements}

We thank the ICCAD anonymous reviewers for their constructive comments and feedback.
This work was supported by the National Science Foundation (Award \#2245531) as well as a Meta Research Award (2022 Towards Trustworthy Products in AR, VR, and Smart Devices RFP).

{
\bibliographystyle{IEEEtranS}
\bibliography{ICCAD}
}

\end{document}